\documentclass[
 reprint,
 floatfix,
superscriptaddress,
 amsmath,amssymb,
 aps,pra,
]{revtex4-2}
\usepackage[version=3]{mhchem}
\usepackage{graphicx}
\usepackage{dcolumn}
\usepackage{bm}
\usepackage{upgreek}
\usepackage{siunitx}
\usepackage{lipsum} 
\usepackage{soul}
\usepackage{braket}
\DeclareMathOperator{\sech}{sech}

\begin{document}
\preprint{APS/123-QED}

\title{Imaging Harmonic Generation of Magnons}

\author{Anthony J. D'Addario}
\affiliation{Department of Physics, Cornell University, Ithaca, New York 14853, USA}

\author{Kwangyul Hu}
\affiliation{Department of Physics and Astronomy, University of Iowa, Iowa City, Iowa 52242, USA}

\author{Maciej W. Olszewski}
\affiliation{Department of Physics, Cornell University, Ithaca, New York 14853, USA}

\author{Daniel C. Ralph}
\affiliation{Department of Physics, Cornell University, Ithaca, New York 14853, USA}
\affiliation{Kavli Institute at Cornell for Nanoscale Science, Ithaca, New York 14853, USA}

\author{Michael E. Flatt\'e}
\affiliation{Department of Physics and Astronomy, University of Iowa, Iowa City, Iowa 52242, USA}
\affiliation{Department of Applied Physics and Science Education, Eindhoven University of Technology,
P.O. Box 513, 5600 MB, Eindhoven, The Netherlands}

\author{Katja C. Nowack}
\affiliation{Kavli Institute at Cornell for Nanoscale Science, Ithaca, New York 14853, USA}
\affiliation{Laboratory of Atomic and Solid State Physics, Cornell University, Ithaca, New York 14853, USA}

\author{Gregory D. Fuchs}
\email{gdf9@cornell.edu}
\affiliation{Kavli Institute at Cornell for Nanoscale Science, Ithaca, New York 14853, USA}
\affiliation{School of Applied and Engineering Physics, Cornell University, Ithaca, New York 14853, USA}


\begin{abstract}

This work combines theory and experiment to examine the mechanisms underlying the harmonic generation of magnons. We develop a nonlinear spin-wave framework that is directly analogous to harmonic generation in nonlinear optics, and combine it with scanning nitrogen-vacancy (NV) center magnetometry to image and quantify magnonic harmonic generation in a Ni$_{81}$Fe$_{19}$/Pt microstripe. Within this framework, the harmonic response arises from nonlinear magnetization dynamics localized at strongly inhomogeneous textures, such as the sample edges and domain walls, that act as anharmonic confining potentials. Scanning probe imaging confirms that the harmonic response is correspondingly nonuniform and concentrated near the sample edges. We measure an expected nonlinear power-law scaling, a systematic shift toward larger wavevector excitations at higher harmonic order, and a spin-selective response indicative of an increasingly chiral harmonic stray field. These results provide a microscopic understanding of magnonic harmonic generation and highlight its potential for engineering nonlinear functionality in magnonic systems.

\end{abstract}

\maketitle

\section{Introduction \label{sec:Intro}}

Harmonic generation is a nonlinear process in which a driven medium produces waves with frequency components at integer multiples of the excitation frequency. In nonlinear optics, harmonic signals arise from driving an electronic system within an anharmonic potential, resulting in nonlinear susceptibilities that depend on the system symmetry. These nonlinear optical processes enable a wide range of experimental and technological applications.  For example, second-harmonic generation (SHG) is widely used for optical frequency doubling of lasers~\cite{Franken_1961}, ultrafast pulse characterization~\cite{Treacy_1969, Kane_1993}, and nonlinear optical microscopy of non-centrosymmetric media~\cite{Campagnola_1999, Moreaux_2000, Campagnola_2003}. Third-harmonic generation (THG), a higher-order harmonic process, is employed as a complementary microscopy tool to study interfacial structures of materials irrespective of inversion symmetry~\cite{Barad_1997, Squier_1998}. In addition, nonlinear frequency-mixing processes such as optical parametric oscillation and four-wave mixing enable widely used tunable laser sources~\cite{Giordmaine_1965}. Together, these applications highlight the broad utility of harmonic generation in driven wave systems.

Magnetic materials provide a natural platform to explore analogous nonlinear phenomena of spin-wave excitations, known as magnons. High-harmonic generation of magnons has recently been observed in soft ferromagnets including Ni$_{80}$Fe$_{20}$ (permalloy), detected using an ensemble of nitrogen-vacancy (NV) centers in nanodiamonds dispersed on the surface of a driven magnetic device~\cite{Koerner_2022, Lan_2025}. The harmonic fields generated by magnons are coherent~\cite{Lan_2025}, which is consistent with the coherence of nonlinear frequency conversion. Both even and odd harmonic orders were observed in these experiments~\cite{Koerner_2022,Lan_2025}, indicating that an underlying nonlinear magnon generation mechanism is not constrained to a single symmetry. However, many central questions remain. Direct spatial imaging of harmonic magnons with high spatial resolution, information about the wavevectors of harmonic magnons, and even confirmation of expected power-law scaling is still lacking, obscuring the underlying mechanism of this phenomenon.

In this work, we combine the results of nanoscale resolution imaging using scanning NV center microscopy with a theoretical framework that accounts for the microscopic mechanism of our observations in close analogy to nonlinear optical harmonic generation. This framework attributes the emergence of harmonic magnons to effective anharmonic potentials localized at magnetization textures such as sample boundaries and domain walls, a prediction directly confirmed by nanoscale imaging. Systematic variation of the drive amplitude demonstrates an expected power-law scaling, confirming the nonlinear origin of the response. High-resolution imaging combined with controlled variation of the NV-to-sample separation also allows us to extract the characteristic wavevectors of the harmonic spin waves. Finally, measurements performed on both NV ground state transitions reveal signatures of chirality, which become increasingly pronounced at higher harmonic order. The combined experimental and theoretical analysis further reveals a magnonic analog of the optical Kerr effect, in which the nonlinear driving modifies the underlying magnetic texture. These results establish the microscopic mechanism of magnon-driven harmonic generation, which may have applications in future magnonics-based information technology.

\section{Background \label{sec:HGIntro}}

Harmonic generation has been extensively studied in nonlinear optics, where it arises from the anharmonic nature of the electronic restoring potential in electronic systems and is commonly described through a power-series expansion of the material polarization in the applied electric field~\cite{Boyd_2020}. One can understand magnon harmonic generation in a similar framework by considering an expansion of the local magnetization vector as a power series in the driving magnetic field in an anharmonic magnon potential, given as
\begin{equation}\label{eq:nonlinear_M(t)}
\mathbf{M}(t) = \mu_0 \left[\chi^{(1)} \mathbf{H}(t) + \chi^{(2)} \mathbf{H}^2(t) + \chi^{(3)} \mathbf{H}^3(t) + \cdots \right],
\end{equation}
where $\mu_0$ is the permeability of free space, $\mathbf{H}(t)$ is the driving field amplitude, $\chi^{(1)}$ is the linear magnetic susceptibility, and the $\chi^{(n)}$ are the $n^{\text{th}}$-order magnetic susceptibilities that describe the nonlinear contributions. In this phenomenological picture, the higher-order terms generate oscillatory magnetization components at integer multiples of the drive frequency, providing a direct magnetic analogue of optical harmonic generation.

Linear micromagnetic theory shows that inhomogeneous magnetization textures act as effective confining potentials for spin waves, supporting localized eigenmodes bound to the texture~\cite{Yan_2011,Jorzick_2002,McMichael_2006}. This effective-potential picture applies broadly to static magnetization configurations with strong spatial gradients. In particular, domain walls and sample boundaries give rise to localized spin-wave modes confined near the underlying texture, as explicitly demonstrated for the geometries considered in Appendix~\ref{App:spinwave_pot}.

Building on this established framework, we perturbatively extend the effective-potential description into the nonlinear regime. Magnetic nonlinearities introduce anharmonic terms into the confining potential, enabling frequency mixing and harmonic generation. Within this picture, the fundamental response is proportional to the driving field amplitude and governed by the linear susceptibility $\chi^{(1)}$, while higher-order susceptibilities $\chi^{(n\ge2)}$ give rise to additional response components at integer multiples of the drive frequency, shown in Appendix~\ref{App:nonlinear_harmonics}. This mechanism is further supported by micromagnetic simulations, which explicitly demonstrate the emergence of harmonic spin-wave excitations localized at domain walls and boundaries, shown in Appendix~\ref{App:Mumax_sims}. Similar texture-localized harmonic responses have been observed in previous NV-based studies of driven permalloy films~\cite{Koerner_2022, Lan_2025}.

Within this framework, several experimental signatures are expected. Because harmonic generation relies on localized eigenmodes supported by magnetization textures, the harmonic response should be spatially concentrated at domain walls, sample boundaries, or other regions of strong magnetization gradients. Local pinning sites and defects can similarly modify the magnetization landscape, giving rise to additional confinement and potentially finer spatial structure in the harmonic response. In addition, the amplitude of the $n^{\mathrm{th}}$ harmonic should scale as a power law in the drive amplitude, as implied by the nonlinear susceptibility expansion in Equation~\ref{eq:nonlinear_M(t)}. In the following sections, we test these predictions using spatially resolved NV magnetometry measurements to directly image harmonic generation in a driven permalloy film.

\section{Results \label{sec:Results}}

\subsection{Spatial Mapping of Harmonic Generation}

\begin{figure}
\centering
\includegraphics[scale=1]{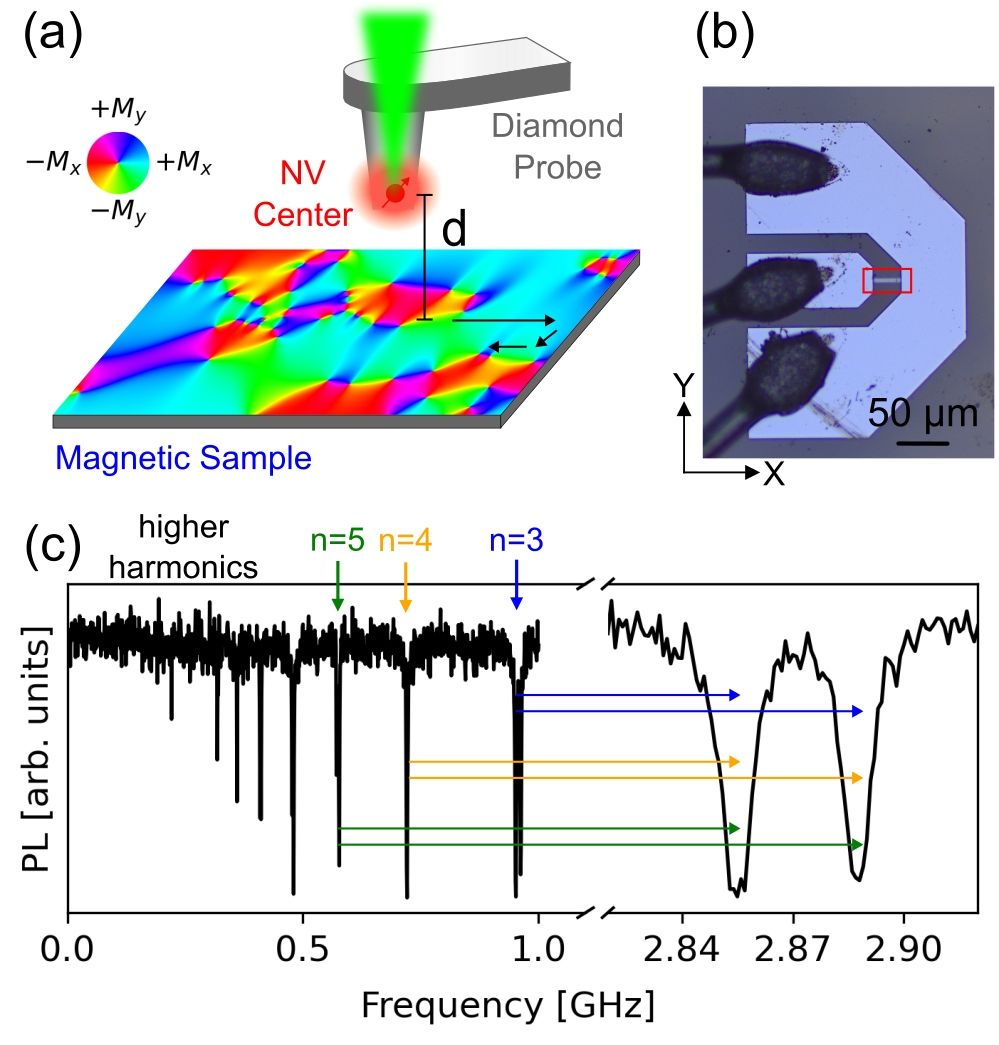}
\caption{{\label{fig:SchematicandDevice}}(a) Schematic of the scanning nitrogen-vacancy (NV) microscope. A single NV center near the apex of a diamond probe detects stray magnetic fields from the sample while raster scanning at a distance $d$. The sample is illustrated with a simulated in-plane magnetization texture. (b) Optical microscope image of the Ni$_{81}$Fe$_{19}$ (5~nm) / Pt (5~nm) device investigated in this work (red box). (c) Optically detected magnetic resonance (ODMR) spectra showing the NV spin resonances near $D \approx 2.87$ GHz and harmonic signals appearing when $n f_0 \approx D$. The low and high frequency spectra are acquired in separate measurements and are stitched together here for clarity.}
\end{figure}

To image harmonic magnon generation, we use a home-built scanning NV microscope, shown in Figure~\ref{fig:SchematicandDevice}(a). A single NV center with an NV-to-sample distance of $d$ measures the projection of the stray magnetic field from our sample along the NV quantization axis with a spatial resolution set by this separation. The probe is mounted on a tuning fork for AFM-based height control.  Figure~\ref{fig:SchematicandDevice}(b) shows an optical microscope image of our $30~\mu\mathrm{m} \times 4~\mu\mathrm{m}$ Ni$_{81}$Fe$_{19}$ ($5~\mathrm{nm}$) / Pt ($5~\mathrm{nm}$) stripe. An AC current is driven directly through the $\sim$160~$\Omega$ device via wire bonds, producing an oscillating Oersted magnetic field that couples to the nearby NV center, while simultaneously driving nonlinear magnetization dynamics in the permalloy through both the Oersted field and spin–orbit torques arising from the Pt layer. All measurements were performed at room temperature.

Figure~\ref{fig:SchematicandDevice}(c) shows optically detected magnetic resonance (ODMR) measurements revealing harmonic generation of magnons detected $50~\mathrm{nm}$ above the sample. During ODMR, the NV spin transitions $\ket{0}\leftrightarrow\ket{\pm1}$ appear as dips in the emitted NV photoluminescence (PL) when the applied microwave frequency matches the NV spin transition frequencies~\cite{Barry_2020}. The right panel shows a standard ODMR spectrum centered about the zero-field splitting $D = 2.87$~GHz, revealing the two ground-state resonance frequencies $f_{\ket{0}\leftrightarrow\ket{-1}}$ and $f_{\ket{0}\leftrightarrow\ket{+1}}$. Because the NV center responds selectively to magnetic fields oscillating near these resonance frequencies, ODMR can be used to detect harmonic stray fields, with the NV acting as a narrowband detector of magnetic fields near $f \approx D$. When the sample is driven at frequency $f_0$, nonlinear magnetization dynamics produce oscillations of the magnetization at integer multiples $n f_0$, resulting in stray magnetic fields at the corresponding harmonic frequencies. Therefore, the $n^{\mathrm{th}}$ harmonic magnon is detected when $n f_0 \approx D$, corresponding to a drive frequency $f_0 \approx D/n$. Consequently, the left panel shows higher-order harmonic signals appearing at progressively lower drive frequencies, $f_{\ket{0}\leftrightarrow\ket{-1}}/n$ and $f_{\ket{0}\leftrightarrow\ket{+1}}/n$, reflecting the stray magnetic field component of the $n^{\mathrm{th}}$ harmonic oscillating at the NV resonance frequency.

\begin{figure}
\centering
\includegraphics[scale=1]{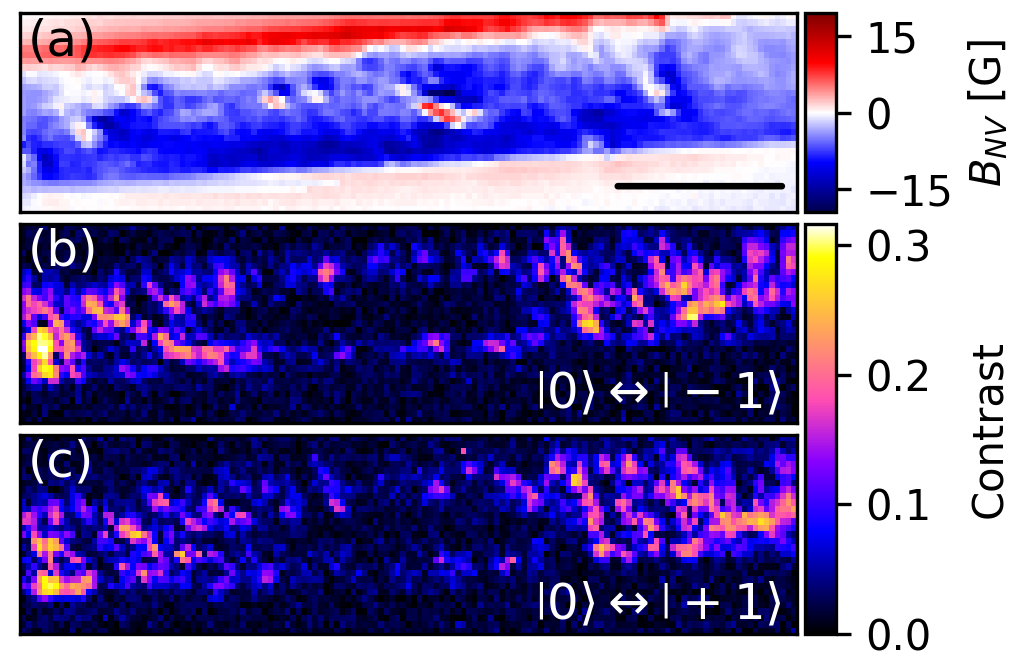}
\caption{{\label{fig:FullStripe}}Results of a wide-field scanning NV measurement. (a) Scanning NV ODMR image showing the stray magnetic field produced by the permalloy stripe under a static magnetic field of $\sim 25~\mathrm{G}$. (b,c) Scanning ODMR measurements of the amplitude of the third-harmonic signal for the $\ket{0} \leftrightarrow \ket{-1}$ and $\ket{0} \leftrightarrow \ket{+1}$ transitions, respectively. The scale bar in (a) is $5~\mu\mathrm{m}$ and applies to all images.}
\end{figure}

To spatially resolve the harmonic generation throughout the permalloy stripe, we first characterize the magnetic stray field. Figure~\ref{fig:FullStripe}(a) shows a wide-field scanning NV measurement of the static stray magnetic field $B_{NV}$ produced by the permalloy magnetization texture projected along the NV quantization axis $u_{NV} = (0,-\sqrt{2},-1)/\sqrt{3}$, at an external bias magnetic field of $\sim25$ G. At each pixel, we acquire an ODMR spectrum centered at $D$ using $\sim0.5~\mathrm{mW}$ of microwave (MW) power with an integration time of $20~\mathrm{ms}$ per frequency point. Appendix~\ref{App:ODMR_fits} describes the Lorentzian fitting procedure used to determine the two ground-state resonance frequencies and extract the local magnetic field strength. The NV center is positioned $200~\mathrm{nm}$ above the sample with a lateral step size of $200~\mathrm{nm}$ between points, corresponding to the expected spatial resolution set by the NV-to-sample separation. This $28~\mu\mathrm{m} \times 6~\mu\mathrm{m}$ scan is performed to image nearly the entire stripe. Both the top and bottom edges of the stripe are clearly visible in this measurement. The observed spatial structure in the stray field indicates that the magnetization is not uniform throughout the sample and that the applied bias field does not fully saturate the permalloy.

Next, we map the third-harmonic magnon signal across the stripe by acquiring an additional ODMR measurement centered at $D/3$ at each scan position using $\sim16~\mathrm{mW}$ of microwave (MW) power, integrating for $20~\mathrm{ms}$ at each frequency point. For a given pixel, the third-harmonic resonances are expected at $f_{\ket{0}\leftrightarrow\ket{-1}}/3$ and $f_{\ket{0}\leftrightarrow\ket{+1}}/3$, where the fundamental transition frequencies $f_{\ket{0}\leftrightarrow\ket{\pm1}}$ are determined from the corresponding ODMR measurement in Figure~\ref{fig:FullStripe}(a). Appendix~\ref{App:ODMR_fits} describes the fitting procedure used to extract the harmonic contrast amplitudes from the ODMR spectra acquired near $D/3$. Across the stripe, the third-harmonic resonance frequencies vary by approximately $80~\mathrm{MHz}$ due to spatial variations in the local magnetization. The resulting third-harmonic contrast maps for the $\ket{0}\leftrightarrow\ket{-1}$ and $\ket{0}\leftrightarrow\ket{+1}$ transitions are shown in Figures~\ref{fig:FullStripe}(b) and (c), respectively.

These results demonstrate that third-harmonic generation is highly localized to regions of strong magnetization inhomogeneity. The harmonic response is strongly concentrated near the sample boundaries, while the interior of the stripe exhibits significantly weaker signal. In particular, harmonic generation is enhanced toward the left and right ends of the stripe. We attribute this asymmetry to the shape anisotropy of the stripe, which favors magnetization alignment along the long $x$ axis. Because the microwave drive field is applied along $\hat{y}$, the magnetization is driven most efficiently when it lies perpendicular to the drive field. This condition is naturally satisfied near the top and bottom edges and toward the ends of the stripe, where the magnetization begins to deviate from uniform alignment. In addition, along the edges, the harmonic contrast exhibits smaller-scale spatial variations, suggesting the presence of additional anharmonic potentials such as local pinning sites. This boundary localization and confinement are consistent with theoretical predictions that magnetic inhomogeneities support enhanced nonlinear response and spin-wave confinement as shown in Appendix~\ref{App:spinwave_pot}.

Overall, these measurements demonstrate that scanning NV magnetometry enables direct, spatially resolved imaging of magnonic harmonic generation in a ferromagnetic thin film. By mapping the harmonic stray fields, we visualize where nonlinear magnetization dynamics are enhanced, revealing strong localization to regions of magnetic confinement in qualitative agreement with our theoretical analysis. Because the magnetization in our permalloy stripe lies predominantly in the film plane with no strong preferred orientation, the stray-field measurement cannot be uniquely inverted to produce magnetic domain image that can be correlated with harmonic generation amplitude~\cite{Casola_2018, Dubois_2022}. Nevertheless, the spatial maps clearly demonstrate that harmonic generation is strongly localized to regions of magnetic inhomogeneity, consistent with nonlinear spin-wave confinement in magnetization textures.

\subsection{Power-Law Scaling of Harmonic Generation}

Having established the spatial distribution of harmonic generation, we next confirm the characteristic power-law scaling expected for a nonlinear magnetic response, as described by Equation~\ref{eq:nonlinear_M(t)}. To do so, we measure the dependence of the harmonic contrast on the applied drive amplitude using scanning NV magnetometry. A series of measurements are performed over a $2~\mu\mathrm{m} \times 2~\mu\mathrm{m}$ region with a $200~\mathrm{nm}$ step size and an NV-to-sample distance of $200~\mathrm{nm}$. The scan region is centered away from the boundaries, towards the right end of the stripe, where we see harmonic generation not just confined to the boundaries. At each spatial position, multiple ODMR measurements are acquired to simultaneously detect the $n=3$, 4, and 5 harmonic responses as a function of the applied root-mean-square (RMS) drive voltage at the same spatial location.

\begin{figure}
\centering
\includegraphics[scale=1]{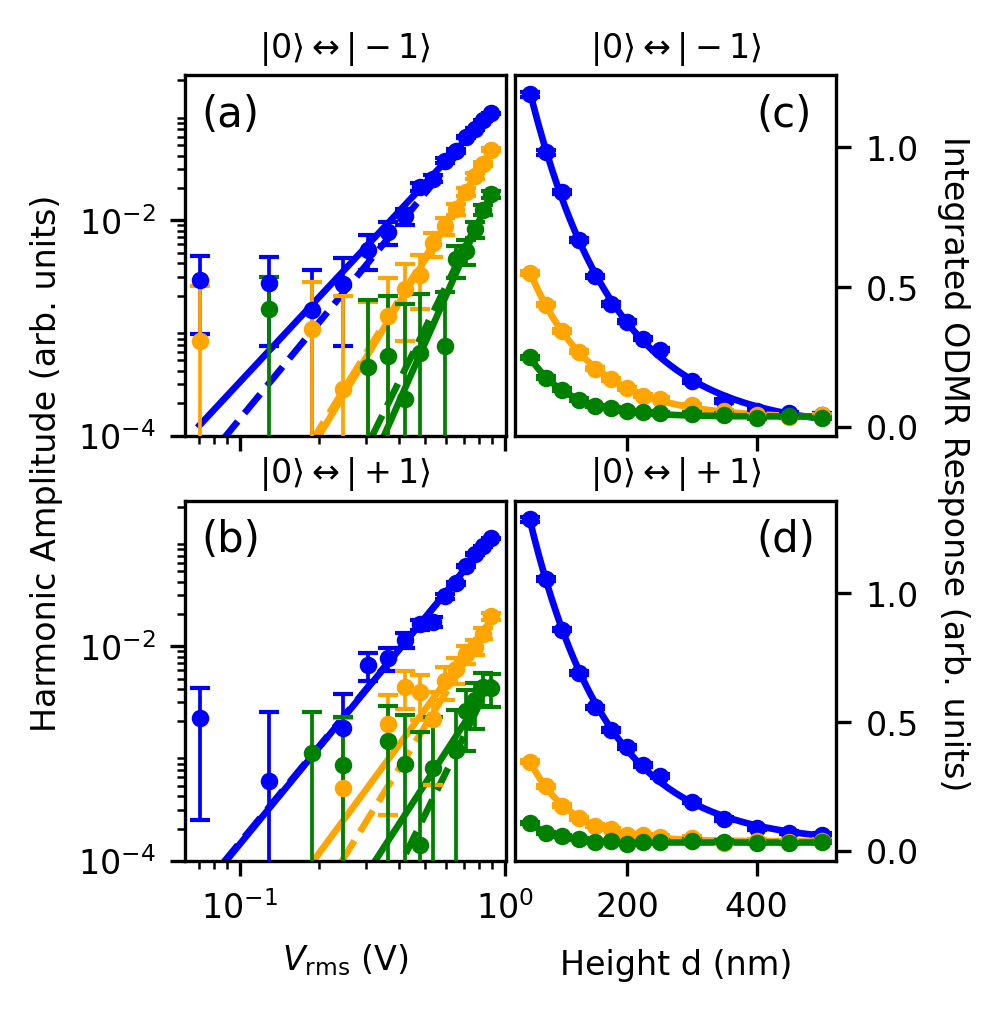}
\caption{{\label{fig:PowerandHeightSeries}}(a,b) Mean harmonic contrast averaged over the scan region as a function of RMS drive voltage $V_{\mathrm{rms}}$ for harmonic orders $n=3$ (blue), $n=4$ (orange), and $n=5$ (green), measured using the $\ket{0}\leftrightarrow\ket{-1}$ (top) and $\ket{0}\leftrightarrow\ket{+1}$ (bottom) NV spin transitions. Dashed lines indicate fixed-exponent power-law trends with $n=3$, 4, and 5, shown as guides to the eye, while solid lines show power-law fits with the exponent treated as a free parameter using Equation~\ref{eq:powerlawfit}. (c,d) Mean harmonic contrast as a function of the NV-to-sample separation $d$, where the colors correspond to the same harmonic orders. Solid lines are fits using Equation~\ref{eq:heightfit}.}
\end{figure}

Figure~\ref{fig:PowerandHeightSeries}(a) and (b) show the mean harmonic contrast averaged over the entire scan region, as a function of the RMS drive voltage for both the $\ket{0} \leftrightarrow \ket{\pm1}$ NV spin transitions. The voltage dependence of the harmonic contrast is modeled using the power law
\begin{equation}\label{eq:powerlawfit}
C(V) = C_0 + a V^n,
\end{equation}
where $a$ is a scaling factor related to the nonlinear magnetic susceptibility $\chi^{(n)}$ in Equation~\ref{eq:nonlinear_M(t)} and $C_0$ is an offset accounting for the noise floor of the ODMR fits, which yields apparent contrast amplitudes of order $\sim5\%$ even in the absence of a detectable harmonic signal due to photon shot noise.

To illustrate the expected $n^{\mathrm{th}}$-order scaling of the harmonic response, we show two fitting approaches. Dashed lines in Figure~\ref{fig:PowerandHeightSeries}(a) and (b) show fixed-exponent power-law trends with $n=3$, 4, and 5, included as guides to the eye to illustrate the ideal nonlinear scaling. Solid lines correspond to fits in which the exponent is treated as a free parameter and extracted directly from the data. To emphasize the power-law behavior on a log-log scale, the fitted background $C_0$ is subtracted from the data prior to plotting. The extracted exponents from these fits are summarized in Table~\ref{tab:power_law_exponents}.

For each harmonic order, the extracted exponent is in reasonable agreement with the expected $n^{\mathrm{th}}$-order scaling, consistent with the nonlinear magnetization model described by Equation~\ref{eq:nonlinear_M(t)}. The free-exponent fits (solid lines) closely track the fixed-exponent guides (dashed lines) shown in Figure~\ref{fig:PowerandHeightSeries}(a) and (b), reinforcing this consistency. Because the harmonic power-law scaling reflects a global nonlinear response of the driven magnetic texture, spatial averaging preserves the voltage exponent even in the presence of local amplitude variations, providing a robust measure of the nonlinear response. This agreement with the expected $n^{\mathrm{th}}$-order scaling provides strong experimental evidence that the observed harmonic signals arise from nonlinear magnetization dynamics.

\begin{table}[t]
    \centering
    \renewcommand{\arraystretch}{1.5}
    \setlength{\tabcolsep}{12pt}
    \caption{Extracted power-law exponents obtained from fits to Equation~\ref{eq:powerlawfit}, where the exponent $n$ is treated as a free parameter, for the data shown in Figure~\ref{fig:PowerandHeightSeries}(a) and (b).}
    \label{tab:power_law_exponents}
    \begin{tabular}{c|cc}
        \hline\hline
        & \multicolumn{2}{c}{$n_{\mathrm{fit}}$} \\
        \cline{2-3}
        & $\lvert 0\rangle \leftrightarrow \lvert -1\rangle$
        & $\lvert 0\rangle \leftrightarrow \lvert +1\rangle$ \\
        \hline
        $n=3$ & $2.64 \pm 0.09$ & $3.02 \pm 0.09$ \\
        $n=4$ & $4.09 \pm 0.27$ & $3.35 \pm 0.60$ \\
        $n=5$ & $5.6 \pm 0.9$   & $3.8 \pm 2.3$ \\
        \hline\hline
    \end{tabular}
\end{table}

While the overall voltage dependence follows the expected nonlinear scaling, systematic deviations arise at the lowest and highest drive amplitudes. At low drive voltages, the harmonic signal approaches the noise floor, leading to increased uncertainty in the extracted scaling. At the highest drive voltages, deviations are most pronounced for the $n=3$ harmonic and attributed to saturation of the NV optical contrast, which limits the measurable signal to approximately $30\%$. Despite these deviations, we see the expected nonlinear behavior where lower-order harmonics exhibit larger signal amplitudes at a given drive voltage, while higher-order harmonics are progressively weaker. This trend reflects the decreasing magnitude of the higher-order nonlinear magnetic susceptibilities $\chi^{(n)}$, indicating that progressively higher-order nonlinear couplings contribute less strongly to the magnetization response.

\subsection{Wavevector Analysis of Harmonic Generation}\label{Sec:heightdep}

Scanning NV magnetometry provides direct sensitivity to the characteristic wavevectors of the harmonic spin-wave excitations through the height dependence of the detected stray ac field. For a spin-wave mode with wavevector $k$, the dipolar stray field decays according to the transfer function $B(k,d) \propto e^{-2kd}$~\cite{Van_der_sar_2015, Du_2017, Purser_2020}. Therefore, by measuring the decay of the harmonic contrast with height, we can extract an effective wavevector characterizing the dominant spatial frequency of the excitation.

We determine this height dependence experimentally by performing a series of scanning NV measurements over a $2~\mu\mathrm{m} \times 2~\mu\mathrm{m}$ region with a $200~\mathrm{nm}$ step size while varying the NV-to-sample distance. Measurements are acquired at two excitation powers of $16~\mathrm{mW}$ and $4~\mathrm{mW}$ in a similar region to the power-dependence series.

Figure~\ref{fig:PowerandHeightSeries}(c) and (d) show the mean harmonic response averaged over the scan region as a function of NV-to-sample separation for both spin transitions at $16~\mathrm{mW}$. The corresponding height-dependent data and fits at $4~\mathrm{mW}$ are shown in Appendix~\ref{App:HeightSeries4mW}. To mitigate systematic variations in the off-resonant photoluminescence with NV-to-sample distance, we quantify the harmonic signal using the integrated ODMR response (contrast $\times$ linewidth) extracted from Lorentzian fits. The height dependence of the harmonic contrast is modeled using
\begin{equation}
C(d) = C_0 + A e^{-2k_{\mathrm{eff}} d},
\label{eq:heightfit}
\end{equation}
where $C_0$ accounts for a constant background contrast offset arising from the ODMR fitting procedure, $A$ is a scaling prefactor that sets the overall magnitude of the harmonic signal, and $k_{\mathrm{eff}}$ is an effective wavevector extracted from the height-dependent decay. Solid lines indicate fits to this model, and the extracted values of $k_{\mathrm{eff}}$ for different harmonic orders, spin transitions, and excitation powers are summarized in Table~\ref{tab:keff_power}.

\begin{table}[t]
    \centering
    \renewcommand{\arraystretch}{1.5}
    \setlength{\tabcolsep}{3pt}
    \caption{Effective wavevectors $k_{\mathrm{eff}}$ extracted from height-dependent fits for different harmonic orders, spin transitions, and excitation powers.}
    \label{tab:keff_power}
    \begin{tabular}{c|cc|cc} 
        \hline\hline
        & \multicolumn{4}{c}{$k_{\mathrm{eff}}$ ($\mathrm{rad}/\mu\mathrm{m}$)} \\
        \cline{2-5}
        & \multicolumn{2}{c|}{16 mW}
        & \multicolumn{2}{c}{4 mW} \\
        \cline{2-5}
        & $\lvert 0\rangle \leftrightarrow \lvert -1\rangle$
        & $\lvert 0\rangle \leftrightarrow \lvert +1\rangle$
        & $\lvert 0\rangle \leftrightarrow \lvert -1\rangle$
        & $\lvert 0\rangle \leftrightarrow \lvert +1\rangle$ \\
        \hline
        $n=3$ & $3.80 \pm 0.05$ & $4.10 \pm 0.06$ & $8.1 \pm 0.4$ & $9.3 \pm 0.5$ \\
        $n=4$ & $5.3 \pm 0.1$   & $7.9 \pm 0.3$   & $12 \pm 2$    & $12 \pm 4$ \\
        $n=5$ & $8.1 \pm 0.4$   & $13 \pm 2$      & ---           & --- \\
        \hline\hline
    \end{tabular}
\end{table}

The effective wavevector $k_{\mathrm{eff}}$ extracted from the height dependence reflects the characteristic spatial scale of the nonlinear magnetic excitations that dominate the measured stray-field signal. In regions of harmonic generation, the magnetization dynamics are governed by anharmonic magnetic potentials whose spatial extent, together with the spin-wave dispersion and excitation efficiency under microwave driving, determines the distribution of accessible wavevectors. Because the NV measurement averages over all excitations within the scan region and filters the signal through the dipolar transfer function, the extracted $k_{\mathrm{eff}}$ represents a weighted characteristic wavevector rather than a single mode~\cite{Van_der_sar_2015, Du_2017, Purser_2020}. We find that $k_{\mathrm{eff}}$ increases systematically with harmonic order, indicating that higher-order harmonics contain progressively greater contributions from larger-$k$ (shorter-wavelength) spin waves generated by the nonlinear magnetization dynamics. In addition, the extracted values of $k_{\mathrm{eff}}$ decrease with increasing excitation power, see Table~\ref{tab:keff_power}, indicating that stronger microwave driving produces harmonic excitations with longer characteristic wavelengths and a more spatially extended nonlinear response.

The systematic increase of $k_{\mathrm{eff}}$ is nontrivial because all harmonic signals are detected at approximately the same frequency, fixed by the NV spin transition near $D = 2.87~\mathrm{GHz}$. In our measurements, the sample is driven at frequencies $f_0 \approx D/n$, such that nonlinear magnetization dynamics generate harmonic oscillations at $n f_0 \approx D$, where they are detected by the NV center. Consequently, although different harmonic orders correspond to different drive frequencies, the stray fields measured by the NV all oscillate near the same detection frequency $f \approx D$. In a simple perturbative picture with a fixed static magnetic potential and an unchanged spin-wave eigenbasis, one would therefore expect similar spatial profiles at this common detection frequency for all harmonic orders. The observed increase of $k_{\mathrm{eff}}$ indicates a breakdown of this simple picture.

To confirm the origin of this behavior, we perform micromagnetic simulations using a domain-wall toy model described in Appendix~\ref{App:Mumax_sims}, building upon previous work studying the nonlinear spin waves in magnetic textures~\cite{Koerner_2022, Lan_2025}. When the system is driven at $f_{\mathrm{drive}} = D/n$ and the demagnetization field is time-averaged at $f = D$, corresponding to the NV detection frequency, the domain wall exhibits distinct spatial profiles for different harmonic orders. These simulations demonstrate that nonlinear driving modifies the magnetization texture itself, generating spin-wave excitations with progressively different characteristic wavevectors as the harmonic order increases, see Appendix~\ref{App:drive_demag}.

This behavior is the magnetic analog of the optical Kerr effect in nonlinear optics, where the refractive index of a material depends on the applied electric field through a nonlinear optical susceptibility~\cite{Boyd_2020}. In magnetic media, the demagnetization field is directly related to the magnetization through the magnetic susceptibility. Under strong microwave driving, the nonlinear response of the magnetization effectively modifies the magnetic susceptibility of the material, leading to harmonic generation accompanied by changes in the spatial structure of the spin wave excitations. This results in the observed increase of the effective wavevector with harmonic order.

\begin{figure*}
\centering
\includegraphics[scale=1]{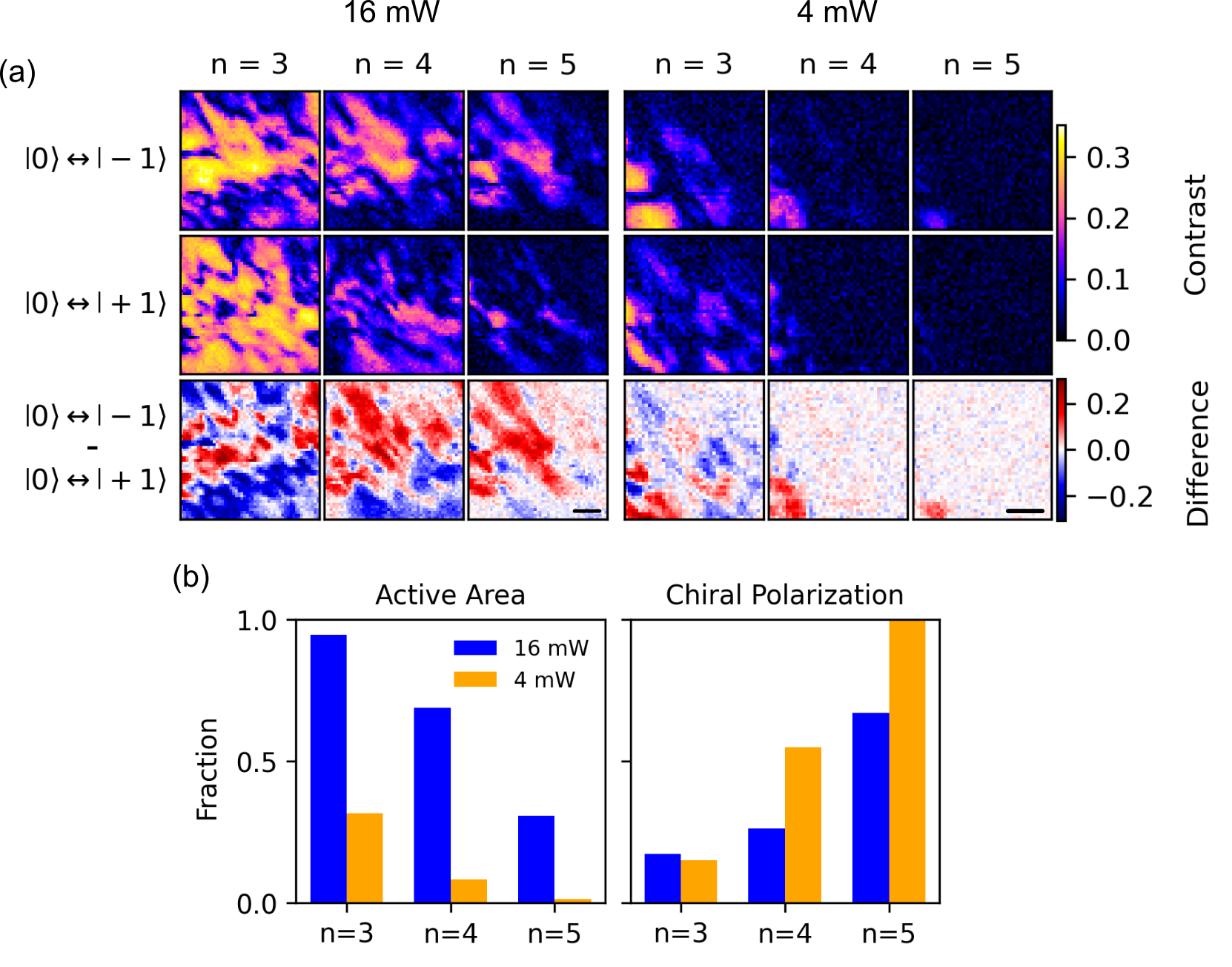}
\caption{{\label{fig:Chirality}}(a) Harmonic contrast maps measured on the $\ket{0} \leftrightarrow \ket{-1}$ (top row) and $\ket{0} \leftrightarrow \ket{+1}$ (middle row) NV spin transitions for harmonic orders $n=3,4,5$. Data are acquired over a $2.5 \times 2.5~\mu\text{m}^2$ region at an excitation power of $16~\mathrm{mW}$ (left) and a $2.0 \times 2.0~\mu\text{m}^2$ region at $4~\mathrm{mW}$ (right). The bottom row shows the pixel-wise difference between the two spin transitions, emphasizing the spin-selective (chiral) nature of the stray magnetic field. Scale bars indicate $1~\mu$m (note that the scan areas differ between the two data sets). (b) Active area and chiral polarization $P_n$ extracted from the harmonic contrast maps, showing a systematic increase in chiral polarization with harmonic order.}
\end{figure*}

In parallel with the height-dependence measurements, we analyze the spatial localization of the harmonic contrast maps to independently extract an effective wavevector. In this approach, the degree of spatial localization reflects the characteristic length scale of the underlying anharmonic magnetic potential responsible for harmonic generation. Therefore, we can directly access the characteristic length scale through analysis of a Fourier-space description of the spatial frequency content of high-resolution harmonic contrast maps.

Figure~\ref{fig:Chirality}(a) shows high resolution scans of the harmonic contrast acquired at a fixed NV-to-sample separation of $\sim 50~\mathrm{nm}$ for excitation powers of $16~\mathrm{mW}$ and $4~\mathrm{mW}$, resolved for both spin transitions $\ket{0}\!\leftrightarrow\!\ket{\pm1}$ and harmonic orders $n=3,4,5$. The scans cover a $2.5\times2.5~\mu\mathrm{m}^2$ region for $16~\mathrm{mW}$ and a $2.0\times2.0~\mu\mathrm{m}^2$ region for $4~\mathrm{mW}$, both sampled with a $50~\mathrm{nm}$ pixel size. A relative lateral drift of approximately $1~\mu\mathrm{m}$ occurs between the two measurements.

From each contrast map in Figure~\ref{fig:Chirality}(a), we compute the two-dimensional Fourier power spectrum $P(\mathbf{k})$ over the scan region, revealing a broad distribution of spatial wavevectors that indicates the presence of multiple spin-wave modes under microwave drive. Because these maps represent the harmonic response of the magnetization under drive, the spatial variations primarily reflect the distribution of dynamically excited spin-wave modes rather than static magnetic domain textures. In this analysis, we interpret the spatial structure as arising from standing spin-wave modes confined by the underlying magnetic texture. We can then define an effective wavevector given by
\begin{equation}
k_{\mathrm{eff}} = \sqrt{\frac{\int k^2 P(\mathbf{k})\, d^2\mathbf{k}}{\int P(\mathbf{k})\, d^2\mathbf{k}}},
\end{equation}
which characterizes the root-mean-square spatial frequency content sensitive to the fine spatial structure within the magnetic texture. The uncertainties are estimated from the variation of $k_{\mathrm{eff}}$ computed over overlapping subregions of each scan, reflecting the spatial variability of wavevectors. The extracted values are summarized in Table~\ref{tab:keff_fft}.

\begin{table}[t]
    \centering
    \renewcommand{\arraystretch}{1.5}
    \setlength{\tabcolsep}{3pt}
    \caption{Effective wavevectors $k_{\mathrm{eff}}$ extracted from the two-dimensional Fourier power spectra of the harmonic contrast maps shown in Figure~\ref{fig:Chirality}, for different harmonic orders, spin transitions, and excitation powers.}
    \label{tab:keff_fft}
    \begin{tabular}{c|cc|cc} 
        \hline\hline
        & \multicolumn{4}{c}{$k_{\mathrm{eff}}$ ($\mathrm{rad}/\mu\mathrm{m}$)} \\
        \cline{2-5}
        & \multicolumn{2}{c|}{16 mW}
        & \multicolumn{2}{c}{4 mW} \\
        \cline{2-5}
        & $\lvert 0\rangle \leftrightarrow \lvert -1\rangle$
        & $\lvert 0\rangle \leftrightarrow \lvert +1\rangle$
        & $\lvert 0\rangle \leftrightarrow \lvert -1\rangle$
        & $\lvert 0\rangle \leftrightarrow \lvert +1\rangle$ \\
        \hline
        $n=3$ & $14.7 \pm 0.8$ & $17.2 \pm 1.2$ & $17 \pm 3$   & $20.9 \pm 2.0$ \\
        $n=4$ & $14.2 \pm 1.2$ & $18.1 \pm 1.3$ & $22 \pm 9$   & $31 \pm 8$ \\
        $n=5$ & $15.4 \pm 1.1$ & $22.9 \pm 2.4$ & $34 \pm 6$   & $48.2 \pm 2.3$ \\
        \hline\hline
    \end{tabular}
\end{table}

The FFT based analysis reveals trends consistent with those observed in the height dependent measurements. We find that the effective wavevector increases with increasing harmonic order and decreases with increasing excitation power, and the relative behavior of the two spin transitions is preserved across both methods. Quantitatively, the wavevectors extracted from height dependent decay are systematically smaller, as large $k$ components of the stray field are exponentially suppressed with increasing NV-to-sample separation by the dipolar transfer function $e^{-2kd}$. In contrast, the high resolution scans performed at a fixed separation of $50~\mathrm{nm}$ are sensitive to a broader range of wavevectors, including higher $k$ components. The two approaches yield consistent measures of the characteristic length scales governing harmonic generation.

\subsection{Chirality of the Harmonic Stray Field}

The harmonic stray magnetic field generated by the driven spin waves exhibits a pronounced chirality. This appears as a clear asymmetry in the harmonic contrast measured on the $\ket{0}\leftrightarrow\ket{\pm1}$ NV spin transitions, as shown in Figure~\ref{fig:Chirality}(a) and across the full permalloy stripe shown in Figures~\ref{fig:FullStripe}(b) and \ref{fig:FullStripe}(c). Example ODMR spectra and fits that illustrate this chiral response are shown in Appendix~\ref{App:ODMR_fits}. As the two spin transitions couple selectively to opposite circularly polarized components of an oscillating magnetic field about the NV quantization axis~\cite{Alegre_2007}, the imbalance between the transition amplitudes therefore directly reflects the chirality of the local AC stray field. In this context, chirality refers to the presence of a harmonic signal on one NV transition and its suppression on the other at the same spatial location. This spin-dependent response is spatially inhomogeneous across the stripe.

To quantify the spatial extent of the chiral response, we define an active region consisting of pixel pairs for which a measurable harmonic contrast is observed in either spin channel. This avoids including large areas of the scan region where no harmonic signal is present, which would artificially bias the analysis toward achiral behavior. Therefore, we apply a $10\%$ contrast threshold motivated by the experimental noise in the ODMR fits, where spectra with no discernible harmonic peak yield apparent contrast amplitudes of approximately $5\%$ due to photon shot noise.

Within this active region, we define a chiral polarization metric
\begin{equation}
P_n = \frac{\left| S_{-1}^{(n)} - S_{+1}^{(n)} \right|}{S_{-1}^{(n)} + S_{+1}^{(n)}},
\end{equation}
where $S_{\pm1}^{(n)}$ denote the spatially integrated harmonic contrast measured on the $\ket{0}\leftrightarrow\ket{\pm1}$ transitions for harmonic order $n$. A value of $P_n=0$ corresponds to equal amplitudes on the two transitions (achiral response), while a finite value reflects an imbalance between the transition amplitudes, indicating net chirality in the field produced by the spin waves. A value of $P_n = 1$ corresponds to complete imbalance between the two transitions. Figure~\ref{fig:Chirality}(b) shows the resulting chiral polarization evaluated over the entire region of interest for all harmonic orders and excitation powers.

We find that while the active area decreases with increasing harmonic order, the chiral polarization $P_n$ increases systematically. Together, these trends indicate that higher-order harmonic generation probes a progressively smaller but more strongly chiral subset of the spin-wave excitations. This behavior is consistent with theoretical predictions that initially achiral spin waves can develop chiral properties through parity mixing between spin-wave modes of different symmetry, even in systems that are inversion symmetric~\cite{Trevillian_2024}. In this framework, modes with larger wavevector are more strongly affected by parity mixing, making higher-$k$ spin waves more susceptible to exhibiting chiral dynamics~\cite{Trevillian_2024}. These observations demonstrate that higher-order harmonic generation provides a sensitive probe of chiral spin-wave dynamics localized to strongly nonlinear regions of the magnetic texture.

\section{Discussion}

In this work, we establish a microscopic framework for magnonic harmonic generation that is directly analogous to harmonic generation in nonlinear optics. In this picture, spatially inhomogeneous magnetization textures act as effective confining potentials for spin waves, while nonlinear magnetization dynamics introduce anharmonic corrections that enable frequency mixing and the generation of higher harmonics. This framework demonstrates that nonlinear driving not only generates higher harmonics confined to the textures, but instead modifies the spatial profile of the underlying magnetization producing spin-wave excitations with progressively larger wavevectors, a magnetic analog to the optical Kerr effect in nonlinear optics. This connects magnonic harmonic generation to familiar concepts from nonlinear optics and nonlinear wave physics more broadly.

By combining this framework with spatially resolved measurements via scanning NV center magnetometry, we both confirm the nonlinear nature of the magnonic harmonic generation and show that it is intrinsically local and highly sensitive to the underlying magnetic texture. The harmonic response is concentrated near regions with strong magnetization gradients, such as the sample boundaries, and exhibits additional localized features consistent with confinement by domain walls or pinning sites. These textures host a spectrum of spin waves with a broad distribution of wavevectors set by the magnetic potential confinement, dispersion, and excitation efficiency. Higher harmonic orders probe progressively shorter length scales (larger wavevectors) and increasingly chiral components of the spin-wave dynamics. 

More broadly, our results identify magnetization textures as natural nanoscale elements for nonlinear frequency conversion in magnetic systems, governed by the nonlinear magnetic susceptibility. This points to the possibility of controlled engineering the nonlinear magnonic responses through patterned boundaries, domain walls, or pinning sites, to tailor the harmonic spatial localization, efficiency, wavevector content, and chirality. The similarity between magnonic and optical harmonic generation suggests that ideas developed in nonlinear optics can be meaningfully translated into the magnonic domain. These insights open new opportunities for investigating nonlinear magnonic functionality and its potential role in future magnon-based information technologies.

\begin{acknowledgments}

We thank Xiaoxi Huang for helpful insights and discussions. Measurements, theory, and data analysis were supported as part of the Center for Energy Efficient Magnonics an Energy Frontier Research Center funded by the U.S. Department of Energy, Office of Science, Basic Energy Sciences at SLAC National Laboratory under contract \#DE-AC02-76SF00515. Device fabrication was performed at the Cornell Nanofabrication Facility (CNF). CNF is a member of the National Nanotechnology Coordinated Infrastructure (NNCI), which is supported by the National Science Foundation (Grant No. NNCI-2025233)

\end{acknowledgments}

\appendix

\section{Spin Wave Excitations in Inhomogeneous Magnetization Textures}\label{App:spinwave_pot}

We begin by establishing the minimal theoretical framework needed to describe spin wave excitations in a spatially varying magnetization texture. Magnetization dynamics are governed by the Landau-Lifshitz (LL) equation,
\begin{equation}
\partial_t \mathbf{m} = -\gamma\,\mathbf{m}\times\mathbf{h}_{\mathrm{eff}},
\end{equation}
where $\gamma$ is the gyromagnetic ratio, $\mathbf{m}$ is the magnetization, and $\mathbf{h}_{\mathrm{eff}}$ is the effective magnetic field. For simplicity, we do not include the damping term. We consider a minimal one-dimensional model for $\mathbf{h}_{\mathrm{eff}}$ including exchange and uniaxial anisotropy, following Reference~\cite{Yan_2011},
\begin{equation}
\mathbf{h}_{\mathrm{eff}} = K m_z \hat{\mathbf{z}} + A\,\partial_z^2 \mathbf{m},
\end{equation}
where $z$ denotes the spatial coordinate, $K$ is the anisotropy constant, and $A$ is the exchange stiffness.

We consider a static domain wall described by the Walker profile as a spatially inhomogeneous magnetization texture,
\begin{equation}
\mathbf{m}_0(z) = \bigl(\sin\theta_0\cos\phi_0,\, \sin\theta_0\sin\phi_0,\, \cos\theta_0 \bigr),
\end{equation}
where the spherical angles $\theta_0(z)$ and $\phi_0$ parameterize the equilibrium magnetization direction. For example, $\phi_0 = 0$ and $\phi_0 = \pi/2$ correspond to Ne\'el- and Bloch-type walls, respectively. The polar angle $\theta_0(z)$ satisfies
\begin{equation}
A\,\frac{d^2\theta_0}{dz^2} - \frac{K}{2}\sin(2\theta_0) = 0,
\end{equation}
with solution
\begin{equation}
\theta_0(z) = 2\arctan\!\left(e^{z/\Delta}\right), \qquad \Delta = \sqrt{\frac{A}{K}},
\end{equation}
where $\Delta$ is the characteristic magnetic exchange length.

To study spin wave excitations about this static domain-wall texture, we consider small dynamical fluctuations about the equilibrium configuration. We introduce an orthonormal basis
$\{\hat{\mathbf{e}}_r, \hat{\mathbf{e}}_\theta, \hat{\mathbf{e}}_\phi\}$
associated with $\mathbf{m}_0$, and write
\begin{equation}
\mathbf{m}(z,t) = \hat{\mathbf{e}}_r + \left[u(z)\hat{\mathbf{e}}_\theta + v(z)\hat{\mathbf{e}}_\phi \right] e^{-i\omega_0 t},
\end{equation}
with $|u|, |v| \ll 1$, where $\omega_0$ is the angular spin wave frequency. Linearizing the LLG equation in $u$ and $v$ yields
\begin{align}\label{eq:LinLLG_u}
-i\omega_0 u &=
A\frac{d^2 v}{dz^2} - K\cos(2\theta_0)\,v,
\\
-i\omega_0 v &= - A\frac{d^2 u}{dz^2} + K\cos(2\theta_0)\,u.
\label{eq:LinLLG_v}
\end{align}
Defining the complex spin wave amplitude
\begin{equation}
\psi(z) = u(z) - i v(z),
\end{equation}
and using the identity
\begin{equation}
\cos(2\theta_0) = 2\,\mathrm{sech}^2\!\left(\frac{z}{\Delta}\right) - 1,
\end{equation}
Equations~\ref{eq:LinLLG_u} and \ref{eq:LinLLG_v} reduce to
\begin{equation}
i\partial_t \psi = \left[-A\frac{d^2}{d\eta^2} + K\left(1 - 2\,\mathrm{sech}^2\eta\right) \right]\psi,\qquad \eta = \frac{z}{\Delta},
\label{eq:SchrEq}
\end{equation}
where $\eta$ is the dimensionless spatial coordinate measured in units of the characteristic magnetic exchange length $\Delta$.

\begin{figure}
\centering
\includegraphics[scale=1]{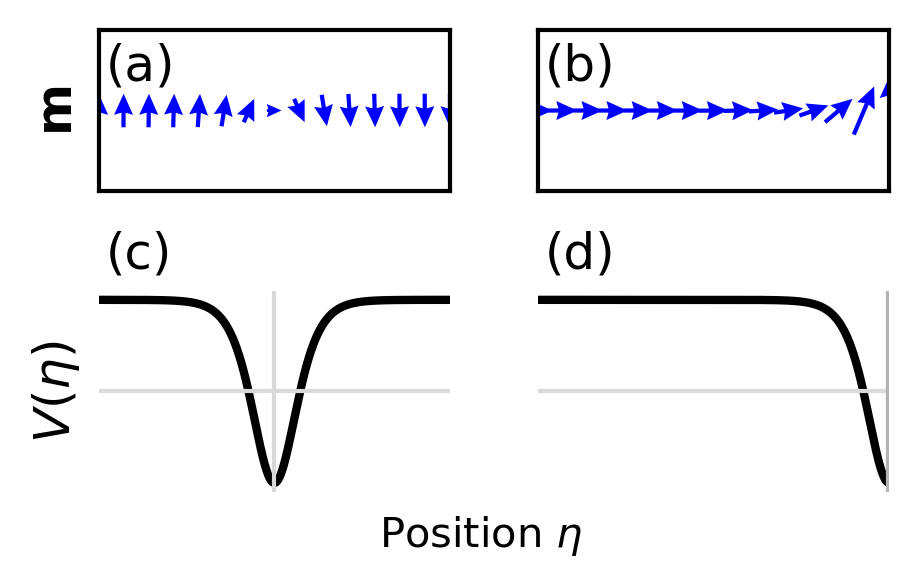}
\caption{{\label{fig:App_SpinConfigurations}}
Spin configurations and corresponding effective magnetic potentials for two representative magnetization textures. (a) Néel-type domain wall described by the Walker profile. (b) Hard boundary geometry, where the domain-wall center is effectively pinned at the boundary. (c,d) Corresponding effective magnetic potentials $\left(1 - 2\,\mathrm{sech}^2\eta\right)$ appearing in Equation~\ref{eq:SchrEq}. In both geometries, the spatial variation of the magnetization produces a localized potential well that can confine spin-wave excitations.}
\end{figure}

Equation~\ref{eq:SchrEq} reveals that a magnetic domain wall acts as an effective confining potential for spin waves. The equation is mathematically equivalent to a one-dimensional Schr\"odinger equation describing a particle in a localized potential well, with the spatial variation of the magnetization as the effective potential term. Although this analysis is strictly linear, it establishes that magnetization textures naturally give rise to localized spin wave eigenmodes bound to the texture.

Importantly, this result is not restricted to domain walls. Any static magnetization texture with strong spatial gradients, including sample edges and boundaries, produces an analogous effective potential. Figures~\ref{fig:App_SpinConfigurations}(a) and (b) show the magnetization configurations for a domain wall and a hard boundary, respectively. The corresponding effective magnetic potentials derived from Eq.~\ref{eq:SchrEq} are shown in Figures~\ref{fig:App_SpinConfigurations}(c) and (d), indicating potential wells that support localized spin wave modes at both domain walls and boundaries.

\section{Higher-Order Harmonics}\label{App:nonlinear_harmonics}

The linear analysis presented in Appendix~\ref{App:spinwave_pot} establishes that inhomogeneous magnetization textures such as domain walls and boundaries support localized spin wave eigenmodes. To describe the emergence of higher-order harmonics, we extend the analysis to the nonlinear regime by including both an external driving field and nonlinear terms arising from the LL equation. The equation of motion for the complex spin wave field $\psi$ can then be written as
\begin{equation}\label{eq:NLLLG_1}
i\partial_t \psi = \mathcal{H}\psi + fe^{-i\omega_d t} + \mathcal{N}(\psi,\psi^*),
\end{equation}
where $fe^{-i\omega_d t}$ represents a monochromatic drive at frequency $\omega_d$, $\mathcal{N}(\psi,\psi^*)$ collects the leading nonlinear contributions, and the linear operator $\mathcal{H}$ is given by
\begin{equation}\label{eq:magneticpotentialwell}
\mathcal{H} = - A\,\frac{d^2}{d\eta^2} + K\left(1 - 2\,\mathrm{sech}^2\eta\right),
\end{equation}
which describes spin wave motion in the effective potential generated by the static magnetization texture.

To analyze the driven nonlinear dynamics, we expand the spin wave field $\psi(\mathbf{r},t)$ in the eigenbasis $\{\Phi_n\}$ of the linear Hamiltonian, defined by $\mathcal{H}\Phi_n = \Omega_n \Phi_n$,
\begin{equation}\label{eq:exp_1}
\psi(\mathbf{r},t) = \sum_n a_n(t)\,\Phi_n(\mathbf{r}),
\end{equation}
where $a_n(t)$ are time-dependent mode amplitudes. Substituting Equation~\ref{eq:exp_1} into Equation~\ref{eq:NLLLG_1} and projecting onto $\Phi_n$ yields
\begin{equation}\label{eq:NLLLG_2}
i\dot{a}_n - \Omega_n a_n + i\Gamma_n a_n = f_n e^{-i\omega_d t} + \left\langle \Phi_n, \mathcal{N}(\psi,\psi^*) \right\rangle,
\end{equation}
where $\Gamma_n$ accounts for the damping of the $n^{th}$ eigenmode and $f_n=\langle\Phi_n,f\rangle$ is the projection of the driving field onto the $n^{th}$ eigenmode.

To analyze the generation of higher-order harmonics, we expand the mode amplitudes in Fourier components at integer multiples of the drive frequency,
\begin{equation}
a_n(t) = \sum_{\ell} a_{n,\ell}\,e^{-i\ell\omega_d t}.
\end{equation}
The nonlinear term can be similarly decomposed as
\begin{equation}\label{eq:NLLLG_3}
\left\langle \Phi_n, \mathcal{N}(\psi,\psi^*) \right\rangle = \sum_{\ell} \left\langle
\Phi_n, g_\ell(a_{n',\ell'}) \right\rangle e^{-i\ell\omega_d t},
\end{equation}
where $g_\ell(a_{n',\ell'})$ denotes nonlinear source terms constructed from products of lower-order harmonic amplitudes with $\ell'<\ell$. This reflects the fact that higher-order harmonics arise through nonlinear mixing of lower-frequency components.
Combining Equations~\ref{eq:NLLLG_2} and \ref{eq:NLLLG_3}, we obtain
\begin{equation}
\left(\Omega_n - \ell\omega_d - i\Gamma_n \right)a_{n,\ell} = f_n\delta_{\ell,1} +
\left\langle \Phi_n, g_\ell(a_{n',\ell'})\right\rangle.
\end{equation}
The resulting harmonic amplitudes are therefore
\begin{equation}
a_{n,\ell}
=
\begin{cases}\label{eq:HHCS}
\dfrac{f_n}{\Omega_n - \omega_d - i\Gamma_n}, & \ell = 1, \\[8pt]
\dfrac{\left\langle\Phi_n,g_\ell(a_{n',\ell'})\right\rangle}{\Omega_n - \ell\omega_d - i\Gamma_n}, & \ell \ge 2.
\end{cases}
\end{equation}

This analysis leads to two important conclusions. First, the fundamental response ($\ell=1$) is resonantly enhanced when the drive frequency matches a linear eigenmode, $\Omega_n=\omega_d$, with an amplitude proportional to the driving field strength. Second, higher-order harmonics ($\ell\ge2$) are resonantly enhanced when $\Omega_n\approx \ell\omega_d$, such that the $\ell^{th}$ harmonic selectively excites eigenmodes whose natural frequencies lie near integer multiples of the drive frequency. Because these eigenmodes are localized within the effective potential well, the resulting harmonic excitations remain spatially confined to the underlying magnetization texture. Therefore, these nonlinear mixing processes generate harmonic responses of the magnetization at integer multiples of the drive frequency.

\section{Magnetization Simulations}\label{App:Mumax_sims}

We perform numerical simulations using the micromagnetic software \textsc{MuMax}$^3$~\cite{Vansteenkiste2014} to investigate magnonic harmonic generation at domain walls and boundaries in a magnetic system representative of our permalloy stripe. The simulation parameters for both the domain-wall and boundary models are summarized in Table~\ref{tab:sim_params}. Wherever possible, we choose material properties, geometry, and driving conditions that closely match our experimental system.

\begin{table*}[t]
    \centering
    \renewcommand{\arraystretch}{1.4}
    \setlength{\tabcolsep}{6pt}
    \caption{Simulation parameters used for the domain wall and boundary models.}
    \label{tab:sim_params}
    \begin{tabular}{c|cc}
        \hline\hline
        \textbf{Parameter} & \textbf{Domain wall} & \textbf{Boundary} \\
        \hline
        \multicolumn{3}{c}{\textit{Mesh and geometry}} \\
        \hline
        Cell size $(C_x,C_y,C_z)$ & $5\times5\times5$ nm$^3$ & $5\times5\times5$ nm$^3$ \\
        Grid $(N_x,N_y,N_z)$ & $400\times400\times1$ & $800\times800\times1$ \\
        Periodic boundary conditions & along $x$ and $y$ & along $y$ \\
        \hline
        \multicolumn{3}{c}{\textit{Material parameters}} \\
        \hline
        Saturation magnetization $M_s$ & $8.6\times10^{5}$ A/m & $8.6\times10^{5}$ A/m \\
        Exchange stiffness $A_\mathrm{ex}$ & $13\times10^{-12}$ J/m & $13\times10^{-12}$ J/m \\
        Gilbert damping $\alpha$ & $0.01$ & $0.01$ \\
        Easy axis $\hat{u}$ & $+\hat{y}$ & $+\hat{y}$ \\
        Anisotropy constant $K_{u1}(y)$
        & $\begin{cases}
        -2.53\times10^{3}~\mathrm{J/m^{3}}, & y>0 \\
        +2.53\times10^{3}~\mathrm{J/m^{3}}, & y<0
        \end{cases}$ & $-2.53\times10^{3}~\mathrm{J/m^{3}}$ \\
        \hline
        \multicolumn{3}{c}{\textit{Magnetic fields}} \\
        \hline
        Static field $\mathbf{B}_\mathrm{stat}$
        & $0.7\hat{x}$ mT
        & $0.7(\hat{x}+\hat{y}+\hat{z})/\sqrt{3}$ mT \\
        AC drive field $\mathbf{B}_\mathrm{ac}(t)$
        & $0.4\sin\bigl(2\pi (2.87~\mathrm{GHz}/3)t\bigr)\hat{y}$ mT
        & $0.4\sin\bigl(2\pi (2.87~\mathrm{GHz}/3)t\bigr)\hat{y}$ mT \\
        \multicolumn{3}{c}{\textit{Time integration}} \\
        \hline
        Total simulation time $s_t$ & $1\times10^{-7}$ s & $1\times10^{-7}$ s \\
        Time step $s_{dt}$ & $1\times10^{-10}$ s & $1\times10^{-10}$ s \\
        \hline\hline
    \end{tabular}
\end{table*}

\begin{figure}
\centering
\includegraphics[scale=1]{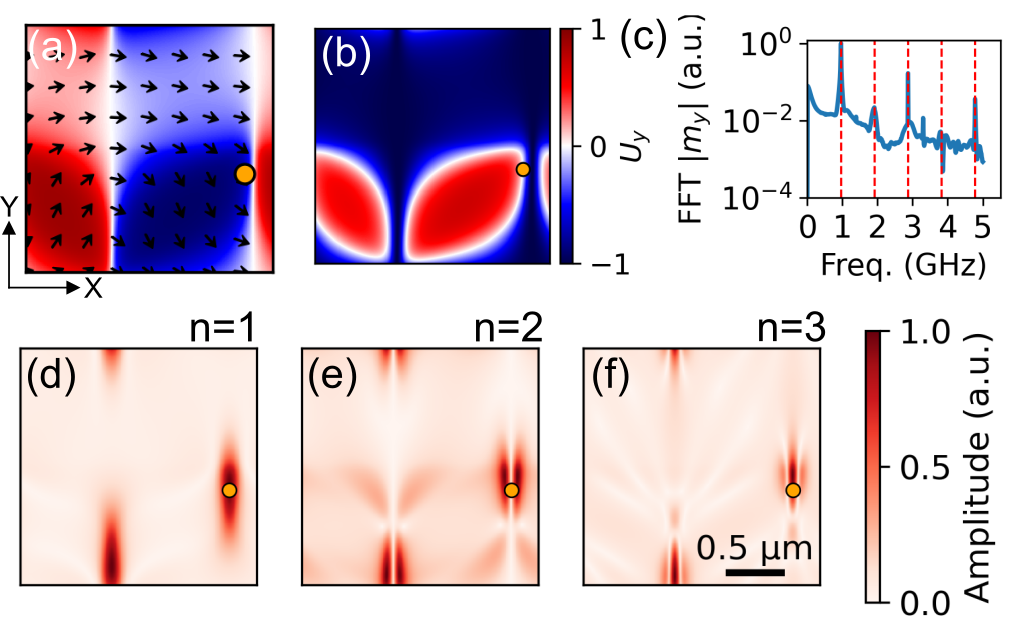}
\caption{{\label{fig:App_DomainWallSim}}Domain wall simulation. (a) Relaxed in-plane magnetization configuration shown as a quiver plot, with color indicating the easy-axis component $m_y$. (b) Effective magnetic potential $U_y$ derived from the static magnetization. (c) FFT of $m_y(t)$ evaluated at the point indicated by the orange marker, showing peaks at integer multiples of the drive frequency. (d–f) Spatial maps of the FFT amplitude of $m_y$ for the first three harmonics ($n=1,2,3$), demonstrating that harmonic generation is strongly localized to the domain-wall regions.}
\end{figure}

Figure~\ref{fig:App_DomainWallSim} illustrates the domain wall simulation. The simulated volume is $2000\times2000\times5~\mathrm{nm}^3$, discretized using a $5\times5\times5~\mathrm{nm}^3$ cell size, matching the thickness of our permalloy stripe. Periodic boundary conditions are applied along the $x$ and $y$ directions to emulate the interior of an extended film and to isolate domain wall dynamics from edge effects. Due to the thin-film geometry, the magnetization remains essentially in-plane.

Next, we describe the material parameters and driving protocol used to generate and excite domain walls in the simulation. We choose typical permalloy material parameter values including the saturation magnetization, exchange stiffness, and Gilbert damping, to perform the simulations \cite{Koerner_2022, Lan_2025}. To stabilize domain walls, we impose a spatially varying uniaxial anisotropy, with opposite signs in the top and bottom halves of the sample. This creates a region where $\hat{y}$ acts as an easy axis and the other as a hard axis. A static magnetic field $\mathbf{B}_\mathrm{stat}$ is applied along $\hat{x}$ to initialize a simplified magnetic configuration containing a small number of domain walls. After relaxation, an oscillating magnetic field $\mathbf{B}_\mathrm{ac}(t)$ is applied along $\hat{y}$ to drive the system. This driving configuration mirrors our experimental geometry, where an AC current flows along the stripe axis and produces an oscillating magnetic field transverse to the easy axis. The drive frequency is chosen as $f = D/3 = 2.87~\mathrm{GHz}/3$, where $D$ is the zero-field splitting of the NV center ground-state spin transition. Consequently, this simulation corresponds to the generation of the $n=3$ harmonic in our experimental measurements.

Figure~\ref{fig:App_DomainWallSim}(a) shows the relaxed in-plane magnetization, where the easy-axis component $m_y$ changes sign across the domain walls and follows the characteristic $\tanh$ profile, $m_y(\eta)\approx\tanh(\eta)$, with $\eta$ denoting the coordinate normal to the wall. Using the identity $\sech^2\eta = 1 - \tanh^2\eta$, the effective magnetic potential well introduced in Equation~\ref{eq:magneticpotentialwell} can be expressed directly in terms of the magnetization as
\begin{equation}
U_y = 1 - 2\sech^2\eta = 2m_y^2 - 1.
\end{equation}
Figure~\ref{fig:App_DomainWallSim}(b) shows this effective potential, highlighting the confining well formed at the domain wall core where $m_y\approx0$.

Figure~\ref{fig:App_DomainWallSim}(c) shows the Fourier spectrum of $m_y(t)$ evaluated at the point indicated by the orange marker, revealing pronounced peaks at integer multiples of the drive frequency. Figures~\ref{fig:App_DomainWallSim}(d)–(f) display the spatial distribution of the FFT amplitude of $m_y$ for the first three harmonics ($n=1,2,3$) across the entire simulated region. In all cases, harmonic generation is strongly localized to the domain-wall regions where $m_y=0$, demonstrating a clear correlation between the domain-wall potential well and the spatial confinement of nonlinear spin wave modes. Higher harmonic orders exhibit additional nodes in the spin wave amplitude, consistent with higher-order standing wave modes bound to the domain wall.

We next apply the same simulation framework to investigate harmonic generation at a sample boundary. In contrast to the domain-wall model, we simulate a larger $4000\times4000\times5~\mathrm{nm}^3$ region with periodic boundary conditions applied only along the $y$ direction, while the system remains finite along $x$. This geometry models a physical boundary within an extended magnetic film, which matches the width of our permalloy stripe.

The material parameters, cell size, and driving conditions are kept identical to those used in the domain-wall simulations to enable a direct comparison. A uniform uniaxial anisotropy is applied throughout the sample with easy axis along $\hat{y}$, and a static magnetic field $\mathbf{B}_\mathrm{stat}$ is applied along the $(\hat{x}+\hat{y}+\hat{z})/\sqrt{3}$ direction, which more closely matches the direction of the static magnetic field in our experimental measurements. As in the domain-wall case, an oscillating magnetic field $\mathbf{B}_\mathrm{ac}(t)$ is applied along $\hat{y}$ at a drive frequency $f = 2.87~\mathrm{GHz}/3$.

\begin{figure}
\centering
\includegraphics[scale=1]{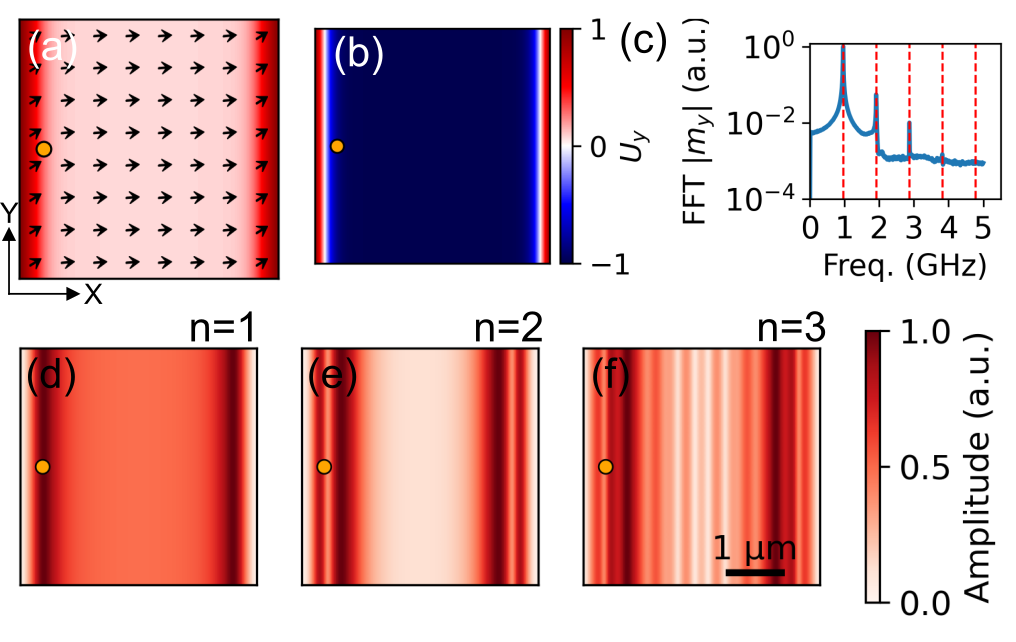}
\caption{{\label{fig:App_BoundarySim}}Boundary simulation. (a) Relaxed in-plane magnetization configuration shown as a quiver plot, with color indicating the easy-axis component $m_y$. (b) Effective magnetic potential $U_y$ derived from the static magnetization. (c) FFT of $m_y(t)$ evaluated at the point indicated by the orange marker, revealing peaks at integer multiples of the drive frequency. (d–f) Spatial maps of the FFT amplitude of $m_y$ for the first three harmonics ($n=1,2,3$), demonstrating that harmonic generation is strongly localized to the boundary region.}
\end{figure}

Figure~\ref{fig:App_BoundarySim}(a) shows that static magnetization profile near the boundary exhibits a spatial variation in the easy-axis component $m_y$, which effectively creates a confining potential analogous to that of a domain wall. As a result, nonlinear spin wave dynamics are preferentially excited near the boundary. Figure~\ref{fig:App_BoundarySim}(b) shows a minimum in the potential energy at the boundaries, and Figure~\ref{fig:App_BoundarySim}(c) confirms harmonic generation from the FFT analysis. Figure~\ref{fig:App_BoundarySim}(d)-(f) shows harmonic generation occurs across the entire boundary. Again, we see behavior consistent with higher-order standing-wave modes at the boundary. However, for increasing harmonic order the corresponding spin wave modes exhibit a broader spatial profile, leading to partial delocalization into the interior of the film. We also see examples of this behavior in our wide-field scanning NV images of the harmonic generation throughout the entire permalloy stripe.

\section{Drive-Dependent Modification of the Demagnetizing Field}\label{App:drive_demag}

Scanning NV magnetometry allows the characteristic spatial scale of the harmonic stray field to be extracted from the height dependence of the measured signal, which we parameterize by an effective wavevector $k_{\mathrm{eff}}$. We find that $k_{\mathrm{eff}}$ increases systematically with harmonic order, even though all harmonics are detected at the same NV resonance frequency $f \approx D = 2.87~\mathrm{GHz}$. To investigate the origin of this behavior, we perform micromagnetic simulations using the domain-wall toy model described in Appendix~\ref{App:Mumax_sims}, driving the system at $f_d = D/n$ and analyzing the magnetization dynamics at the detection frequency $D$.

\begin{figure*}
\centering
\includegraphics[scale=1]{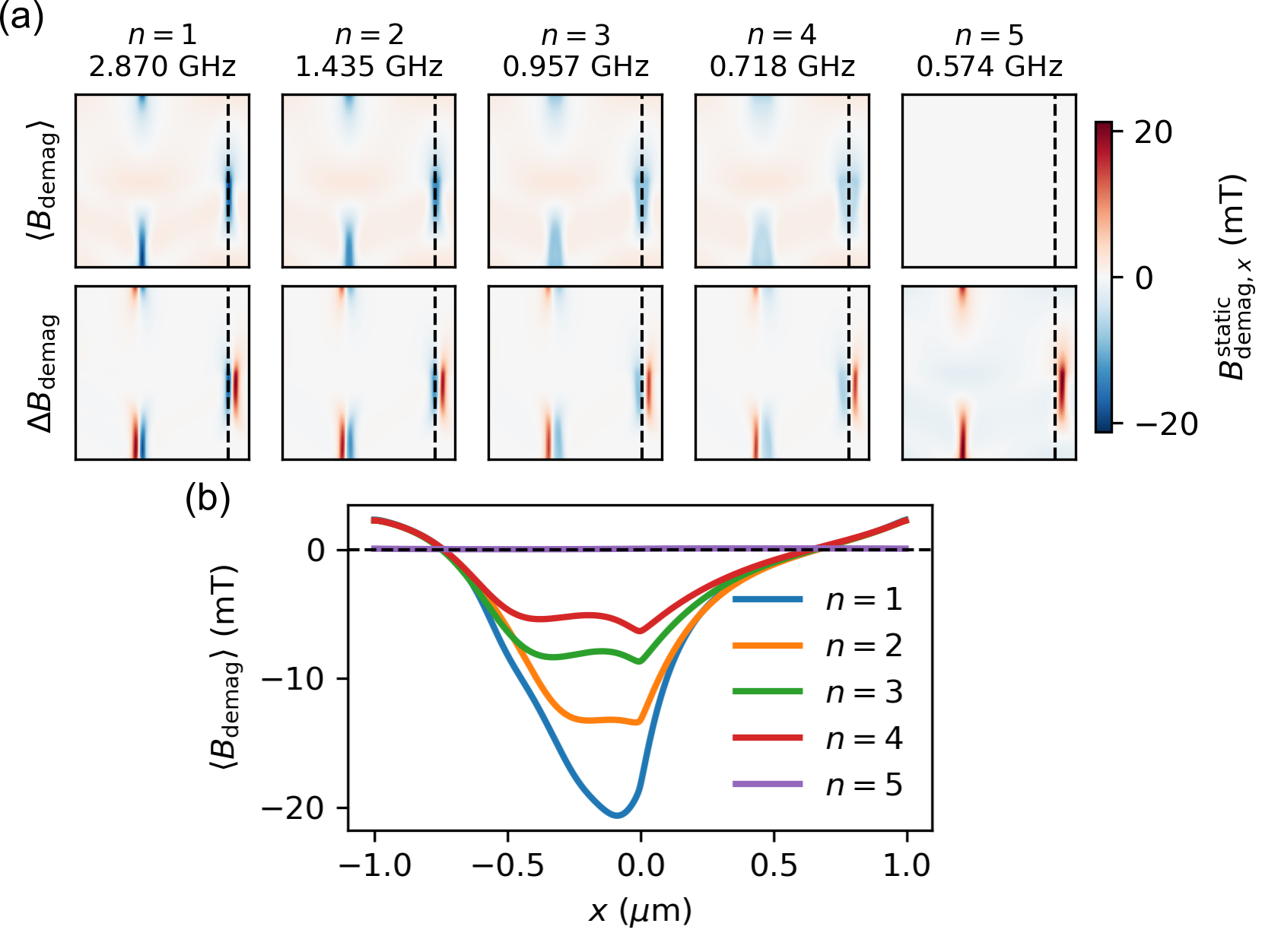}
\caption{\label{fig:App_HarmonicDriving} Demagnetizing field inside the permalloy under driving of various harmonic frequencies. (a) Spatial maps of the demagnetizing field component $B_{\mathrm{demag},x}$ for five driving frequencies $f_d = 2.87/n~\mathrm{GHz}$ corresponding to harmonic orders $n=1$–$5$ (columns). The top row shows the demagnetizing field averaged over the driven steady state, $\langle B_{\mathrm{demag},x} \rangle$, while the bottom row shows the change relative to equilibrium, $\Delta B_{\mathrm{demag},x} = \langle B_{\mathrm{demag},x} \rangle - B_{\mathrm{demag},x}(t{=}0)$. The dashed vertical line indicates the position of the line cut used in (b). (b) Line cuts of $\langle B_{\mathrm{demag},x} \rangle$ taken along the dashed line in (a), illustrating the changing the drive-induced DC response with increasing harmonic order.}
\end{figure*}

Figure~\ref{fig:App_HarmonicDriving}(a) shows the $x$ component of the time-averaged steady-state demagnetizing field, $\langle B_{\mathrm{demag},x} \rangle$, for five different drive frequencies (columns). Although each simulation begins from the same relaxed magnetic configuration, the resulting time-averaged demagnetizing field exhibits a systematic dependence on the drive frequency. This effect is highlighted in the second row of Fig.~\ref{fig:App_HarmonicDriving}(a), which shows the change in the demagnetizing field relative to equilibrium, $\Delta B_{\mathrm{demag},x} = \langle B_{\mathrm{demag},x} \rangle - B_{\mathrm{demag},x}(t=0)$. These maps reveal that strong driving induces a drive-dependent DC modification of the internal magnetic field. Figure~\ref{fig:App_HarmonicDriving}(b) presents line cuts of $\langle B_{\mathrm{demag},x} \rangle$ taken along the dashed lines in (a), demonstrating that both the depth and the shape of the demagnetizing field minimum near the domain wall evolve systematically with increasing harmonic order.

These drive-induced modifications of the static demagnetizing field alter the effective spin-wave potential experienced by the system and thereby modify the spatial structure of the spin-wave modes contributing to the signal at $D$. In particular, changes in the depth and width of the effective magnetic potential modify the spatial confinement of the modes. Stronger confinement produces more localized excitations with larger characteristic wavevectors $k_{\mathrm{eff}}$, despite the fixed detection frequency. Therefore, strong microwave driving dynamically modifies the magnetic susceptibility of the texture through its nonlinear response, analogous to the Kerr effect in nonlinear optics, where the refractive index depends on the applied field.

\section{Extraction of Magnetic Field and Harmonic Contrast from ODMR Spectra}\label{App:ODMR_fits}

At each spatial location in a scanning NV measurement, we acquire an optically detected magnetic resonance (ODMR) spectrum. These spectra are used to determine the local magnetic field by measuring the NV spin transition frequencies near the zero-field splitting $D$. Each ODMR spectrum is fit to the sum of two Lorentzian dips corresponding to the $\ket{0}\leftrightarrow\ket{\pm1}$ transitions,
\begin{equation}\label{eq:ODMR_fit_func}
\mathrm{PL}(f) = \mathrm{PL}_0 - \sum_{i=\pm1} 
\frac{A_i}{1 + \left(\frac{f-f_i}{\Gamma_i}\right)^2},
\end{equation}
where $A_i$ is the amplitude, $f_i$ is the resonance frequency, and $\Gamma_i$ is the linewidth of each transition. The fitted resonance frequencies $f_{\ket{0}\leftrightarrow\ket{\pm1}}$ are used to determine the local magnetic field via the Zeeman splitting,
\begin{equation}
B_{NV} = \frac{f_{\ket{0}\leftrightarrow\ket{+1}} - f_{\ket{0}\leftrightarrow\ket{-1}}}{2\gamma_e},
\end{equation}
where $\gamma_e = 2.8~\mathrm{MHz/G}$ is the electron gyromagnetic ratio~\cite{Barry_2020}.

\begin{figure}
\centering
\includegraphics[scale=1]{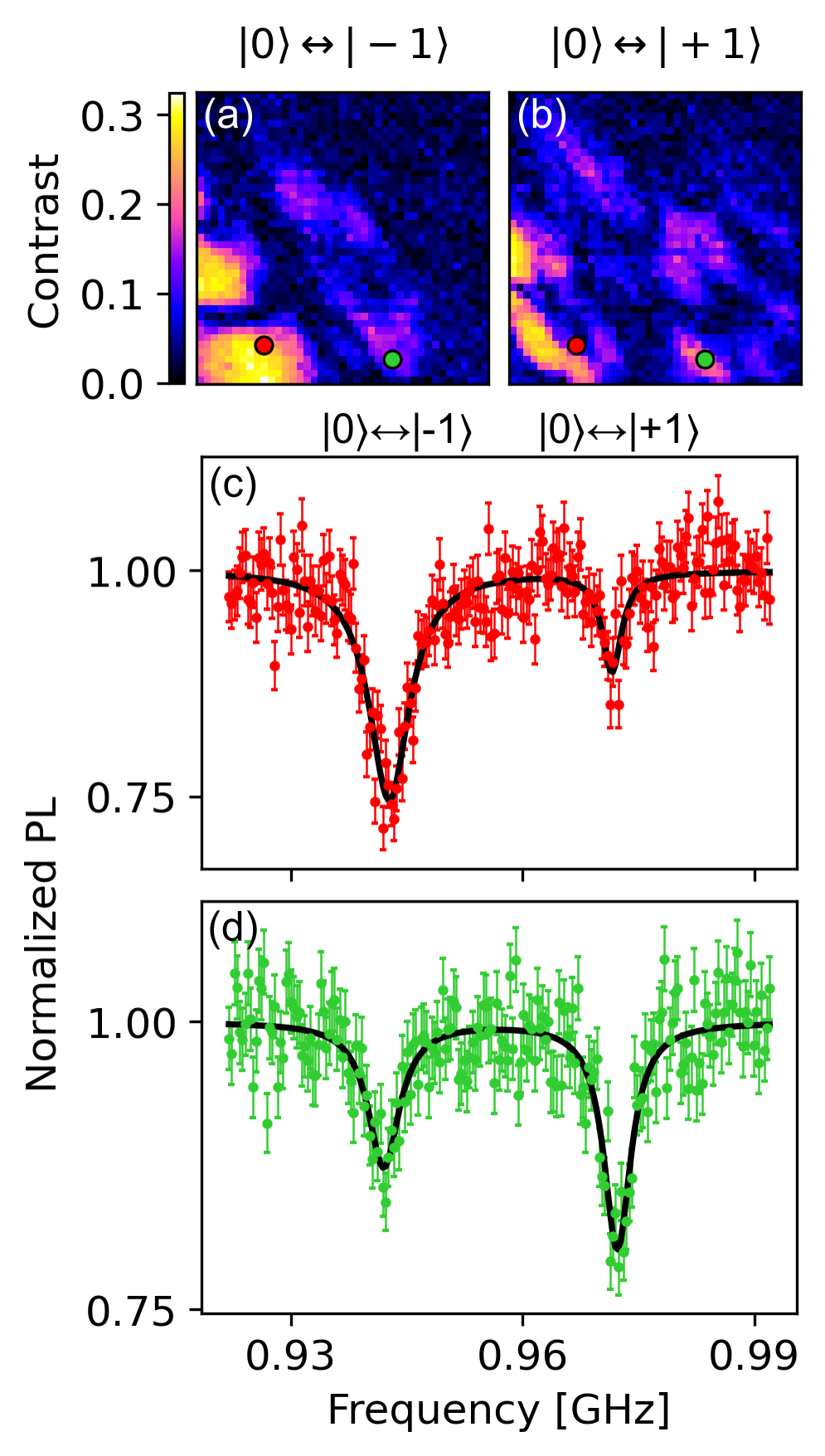}
\caption{{\label{fig:App_ODMR_fitting}}Example of the ODMR fitting procedure used to extract harmonic contrast maps. (a,b) Harmonic contrast maps for the $\ket{0}\leftrightarrow\ket{-1}$ and $\ket{0}\leftrightarrow\ket{+1}$ NV spin transitions, respectively. The colored points indicate the spatial locations from which the ODMR spectra shown below are taken. (c,d) ODMR spectra measured at the corresponding locations for the $\ket{0}\leftrightarrow\ket{-1}$ and $\ket{0}\leftrightarrow\ket{+1}$ transitions. The data are fit to the sum of two Lorentzian dips (black curves) corresponding to the expected harmonic resonance frequencies. The fitted contrast amplitudes are used to construct the spatial contrast maps in (a) and (b).}
\end{figure}

These transition frequencies also determine the expected harmonic detection frequencies $f_{\ket{0}\leftrightarrow\ket{\pm1}}/n$ for harmonic order $n$. ODMR spectra acquired near these frequencies are fit using the same procedure to extract the harmonic amplitudes $A_{\pm1}$. The harmonic contrast reported in the spatial maps corresponds to the normalized amplitude $A_i/\mathrm{PL}_0$. Figures~\ref{fig:App_ODMR_fitting}(a) and (b) show the harmonic contrast maps corresponding to the $n=3$ harmonic in the $4~\mathrm{mW}$ dataset from Figure~\ref{fig:Chirality}(a). Each pixel in these maps corresponds to an ODMR spectrum that is fit independently using Equation~\ref{eq:ODMR_fit_func}. Example spectra and fits from selected spatial locations are shown in Figures~\ref{fig:App_ODMR_fitting}(c) and (d).

The chirality of the harmonic stray field is directly visible in the fitted spectra. Because the two NV spin transitions couple selectively to opposite circularly polarized components of the oscillating magnetic field about the NV quantization axis, differences between the extracted amplitudes $A_{-1}$ and $A_{+1}$ directly measure the chirality of the harmonic magnetic field.

\section{Height-Dependent Wavevector Extraction at Reduced Power}\label{App:HeightSeries4mW}

\begin{figure}
\centering
\includegraphics[scale=1]{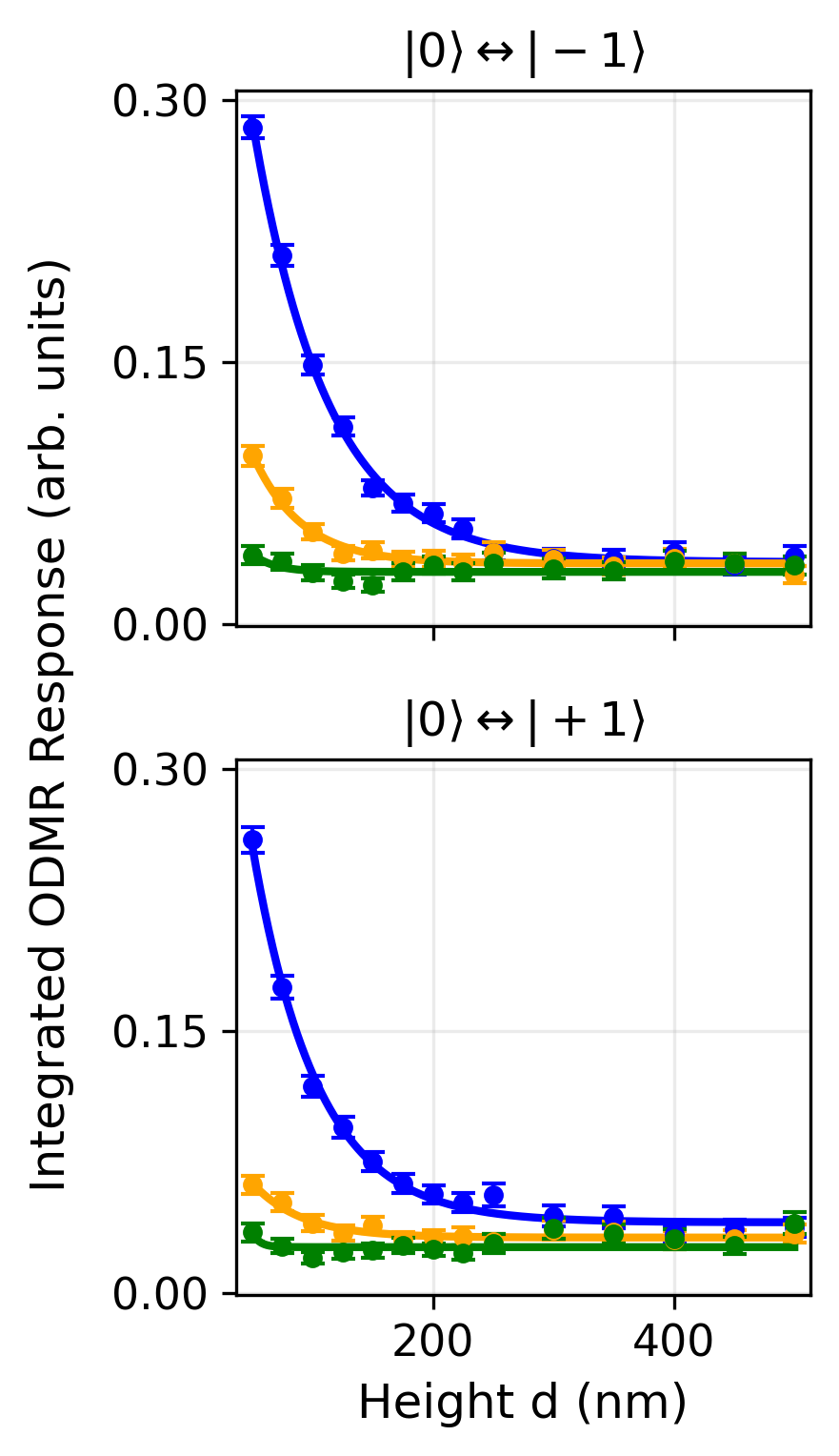}
\caption{{\label{fig:App_HeightSeries4mW}}Mean harmonic contrast as a function of the height of the NV center with respect to the permalloy at $4~\text{mW}$ of microwave power for harmonic orders $n=3$ (blue), $n=4$ (orange), and $n=5$ (green), measured using the $\ket{0} \leftrightarrow \ket{-1}$ (top) and $\ket{0} \leftrightarrow \ket{+1}$ (bottom) NV spin transitions. Solid lines are fits using Equation~\ref{eq:heightfit}.}
\end{figure}

We repeat the height-dependent measurements at a reduced microwave power of $4~\mathrm{mW}$. This dataset is acquired in the same physical location and with the same scan parameters as the $16~\mathrm{mW}$ measurements described in Section~\ref{Sec:heightdep}. The mean harmonic response averaged over the scan region is shown for harmonic orders $n=3$, 4, and 5 and for both NV spin transitions in Figure~\ref{fig:App_HeightSeries4mW}. 

The extracted effective wavevectors at $4~\mathrm{mW}$ follow the same systematic increase with harmonic order observed at $16~\mathrm{mW}$, demonstrating that the trend is robust against changes in excitation power. The data are fit using Equation~\ref{eq:heightfit}, following the same procedure described in the main text. At the reduced excitation power, no saturation of the NV contrast is observed, in contrast to the $16~\mathrm{mW}$ dataset. However, the overall harmonic signal amplitude is significantly smaller, resulting in a more spatially localized response and correspondingly larger extracted values of $k_{\mathrm{eff}}$. For $n=5$, the harmonic signal is too weak to reliably extract a meaningful value of $k_{\mathrm{eff}}$. The fitted parameters are included in Table~\ref{tab:keff_power}.

\clearpage
\bibliography{bibliography}

\end{document}